\documentclass[12pt,preprint]{aastex}

\usepackage{bm}
\usepackage{calc}
\newcounter{address}
\newcommand{\equationname}{equation}
\newcommand{\latin}[1]{\emph{#1}}
\newcommand{\etal}{\latin{et\,al.}}
\newcommand{\ie}{\latin{i.e.}}
\newcommand{\apriori}{\latin{a priori}}

\newcommand{\sn}{{[s/n]}}
\newcommand{\sntotal}{\sn_{\mathrm{total}}}
\newcommand{\fwhm}{{\theta_{\mathrm{FWHM}}}}
\newcommand{\RA}{{\mathrm{RA}}}
\newcommand{\Dec}{{\mathrm{Dec}}}
\newcommand{\vecmu}{\bm{\vec{\mu}}}
\newcommand{\mualpha}{{\mu_\alpha}}
\newcommand{\mudelta}{{\mu_\delta}}
\newcommand{\var}[1]{\mathrm{Var}({#1})}
\newcommand{\unit}[1]{\mathrm{#1}}
\renewcommand{\mag}{\unit{mag}}
\newcommand{\s}{\unit{s}}
\newcommand{\yr}{\unit{yr}}
\newcommand{\km}{\unit{km}}
\newcommand{\pc}{\unit{pc}}
\newcommand{\kpc}{\unit{kpc}}
\newcommand{\mas}{\unit{mas}}
\newcommand{\pix}{\unit{pix}}
\newcommand{\kmpers}{\km\,\s^{-1}}
\newcommand{\masperyr}{\mas\,\yr^{-1}}
\renewcommand{\arcsec}{\unit{arcsec}}
\newcommand{\arcsecperyr}{\arcsec\,\yr^{-1}}
\newcommand{\code}[1]{\texttt{#1}}
\newcommand{\mean}[1]{\left<{#1}\right>}

\newlength{\examplefigurewidth}
\begin{document}
\title{
  Measuring the undetectable:\\
  Proper motions and parallaxes of very faint sources
}

\author{
  Dustin~Lang\altaffilmark{\ref{Toronto}},
  David~W.~Hogg\altaffilmark{\ref{CCPP},\ref{MPIA},\ref{email}},
  Sebastian~Jester\altaffilmark{\ref{MPIA}},
  Hans-Walter~Rix\altaffilmark{\ref{MPIA}}
}

\setcounter{address}{1}
\altaffiltext{\theaddress}{\stepcounter{address}\label{Toronto}
Department of Computer Science, University of Toronto, 6 King's
College Road, Toronto, Ontario, M5S~3G4 Canada}
\altaffiltext{\theaddress}{\stepcounter{address}\label{CCPP} Center
for Cosmology and Particle Physics, Department of Physics, New York
University, 4 Washington Place, New York, NY 10003}
\altaffiltext{\theaddress}{\stepcounter{address}\label{MPIA}
Max-Planck-Institut f\"ur Astronomie, K\"onigstuhl 17,
D-69117 Heidelberg, Germany}
\altaffiltext{\theaddress}{\stepcounter{address}\label{email} To whom
correspondence should be addressed: \texttt{david.hogg@nyu.edu}}

\begin{abstract}
  The near future of astrophysics involves many large solid-angle,
  multi-epoch, multi-band imaging surveys.  These surveys will, at
  their faint limits, have data on large numbers of sources that are
  too faint to be detected at any individual epoch.  Here we show that
  it is possible to measure in multi-epoch data not only the fluxes
  and positions, but also the parallaxes and proper motions of sources
  that are too faint to be detected at any individual epoch.  The
  method involves fitting a model of a moving point source
  simultaneously to all imaging, taking account of the noise and
  point-spread function in each image.  By this method it is possible
  to measure the proper motion of a point source with an uncertainty
  close to the minimum possible uncertainty given the information in
  the data, which is limited by the point-spread function, the
  distribution of observation times (epochs), and the total
  signal-to-noise in the combined data.  We demonstrate our technique
  on multi-epoch Sloan Digital Sky Survey (SDSS) imaging of the SDSS
  Southern Stripe.  We show that with our new technique we can use
  proper motions to distinguish very red brown dwarfs from very
  high-redshift quasars in these SDSS data, for objects that are
  inaccessible to traditional techniques, and with better fidelity
  than by multi-band imaging alone.  We re-discover all 10 known brown
  dwarfs in our sample and present 9 new candidate brown dwarfs,
  identified on the basis of significant proper motion.
\end{abstract}


\section{Introduction}

There are many multi-epoch imaging surveys in progress or coming up,
which will, among other things, deepen our image of the sky and
provide information on source variability and proper motions.  These
surveys include the SDSS Southern Stripe \citep{sdssdr7}, the Dark
Energy Survey, PanSTARRS, LSST, and SNAP.  These surveys promise
proper-motion precisions for well-detected sources on the order of
$\masperyr$ over large parts of the sky.  For context, a typical halo
star at a distance of $10~\kpc$ moving at a transverse heliocentric
speed of $100~\kmpers$ has a proper motion of $2~\masperyr$, and a
typical disk star at $100~\pc$ and $10~\kmpers$ has a proper motion of
$20~\masperyr$.  These surveys therefore have the capability of
revolutionizing our view of the Galaxy and of the Solar neighborhood.

In most conceptions of a proper-motion measurement, one imagines
measuring the position of a source in each of several images, taken at
different times.  A linear trajectory is fitted to the positions,
relative to some reference frame or set of fixed sources or sources
with well measured proper motions.  In its most straightforward form,
this method only works for sources bright enough to be detected
independently at every epoch---or at least most epochs.  In a
multi-epoch survey like the SDSS Southern Stripe, which has $\sim 70$
epochs \citep{sdssdr7}, this limits the sources with measured proper
motions to a small subset of all sources detectable in the combined
data, since the combined data reach $\sim 2.3~\mag$ fainter than any
individual epoch; for typical source populations this represents
increases in population size by factors of $5$ to $25$ at any given
signal-to-noise threshold.  In this paper we present a methodology for
measuring in multi-epoch imaging the proper motions of sources too
faint to detect at any individual epoch.

There are several different technical regimes for these faint-source
proper-motion measurements.  In the ``easy'' regime, the sources of
interest move a distance smaller than or comparable to the
point-spread function width over the duration of the multi-epoch
survey.  In this regime, the sources are easy to detect in the
co-added image, even without taking account of their proper motions;
proper motions can be determined from processing the individual epoch
images after detection in the co-added image.  There is a
``difficult'' regime in which the sources of interest move
substantially more than the width of the point-spread function over
the duration of the survey.  In this regime, the source will not
appear at high significance in the co-added image if it does not
appear at high significance at any epoch, because its different
appearances in the different individual-epoch images do not overlap.
In principle, the difficult regime can be addressed by brute force
with large computing resources.  In the context of outer Solar-System
bodies, brute-force search in the narrow range of expected motions is
feasible (for example, \citealt{bernstein04a, fuentes2008}).  In this
paper, we consider only the easy regime.

\paragraph{Modeling the data:}
The traditional method for measuring a stellar proper motion with a
set of images taken at different times is as follows: Detect the star
at each observed epoch; measure its centroid (by, for example, finding
the peak or first moment of the flux) at each observed epoch; and fit
a linear motion to the measured positions and times.  This procedure
obtains a proper motion, but it puts an unnecessary requirement on the
data: that the star be detectable at every epoch.  It also puts an
unnecessary burden on the data analyst: it requires decision making
about detection and centroiding of the stars at each epoch, decisions
that matter at low signal-to-noise, or when faced with data issues
such as bad pixels or strong variations in noise from pixel to pixel.

Our new approach is to \emph{model} all individual-epoch images
simultaneously with a single point source that is permitted to have a
non-zero parallax and proper motion.  This approach combines the
individual-image positional measurement and the determination of the
parallax and proper motion, and determines all of these simultaneously
by making a statistically ``good'' model of the union of all the data.

In any well-understood imaging survey, each image will have a
per-pixel noise model, photometric calibration parameters, and a model
of the point-spread function.  In any sufficiently small patch of the
sky, if the foreground-subtracted intensity in that patch is dominated
by a small number of point sources, it is possible to make an accurate
model of all of the pixels in the data set that contribute signal to
that small patch.  In this model of the patch, the fluxes, angular
positions, parallaxes, and proper motions of the stars in the patch
are simply parameter values in the well-fitted models.  In other
words, we are assuming that it is possible to model the set of pixels
(from all of the images) that contribute to the patch with a
$6N$-dimensional model that consists of a set of $N$ moving point
sources.

The proper motions determined by image modeling have several
advantages over those determined by the traditional method: They
require fewer decisions about measurement techniques (although they do
require a good model of the data, including point-spread function);
they use all of the information in all of the pixels, not just those
pixels involved in traditional centroiding; they gracefully handle
missing data due to bad pixels or cosmic rays (assuming the bad pixels
have been flagged); they require the investigator to make explicit the
assumptions about the physical properties of the image and the noise;
they can be made to properly propagate pixel-value uncertainties into
parameter uncertainties (in this case, proper motion uncertainties);
they are the result of optimization of a well-justified scalar
objective function (in this case the likelihood). Most importantly for
what follows, they can be determined in data sets in which the stars
are not well detected at any individual epoch, but only appear in the
\emph{combination} of the images.  In a data set with $\sim 70$ similar
epochs (such as the SDSS Southern Stripe), this corresponds to an
increase in the number of available targets by factors of $5$ to $25$
(assuming source populations double to quadruple with each magnitude
of depth).

Here we propose, build, test, and use an image-modeling system for the
determination of stellar proper motions.  We show that it can work
down to low signal-to-noise ratios and that it makes measurements in
real data that fully exploit the information available.  We also use
it to discover interesting new astrophysical sources.  An
approximation to the technique used here has been used previously in
the Solar System literature \citep{bernstein04a}.

\paragraph{Proper-motion and parallax uncertainties:}
Consider a well-sampled image $i$ with a point-spread function of full
width at half maximum $\fwhm_i$.  The signal-to-noise at which the
flux of a point source can be measured, $\sn_i$, is the sum in
quadrature of the signal-to-noise contributions from pixels within the
point-spread function.  A point source measured with signal-to-noise
$\sn_i$ in a single image can be centroided with (RMS) uncertainty
$\sigma_{\theta,i}$ of
\begin{equation}
  \sigma_{\theta,i} \approx \frac{\fwhm_i}{\sn_i} \quad ;
\end{equation}
details such as the shape of the point-spread function introduce
factors of order unity \citep{king1983}.

If we have $N$ such images spanning some time interval, we might hope
to obtain a proper motion estimate with uncertainty $\sigma_{\mu}$
limited by the point-spread function, the time interval, and the total
signal-to-noise
\begin{equation}
  \sntotal^2=\sum_{\mathrm{images}\ i}\sn_i^2
  \label{eq:sntotal}
\end{equation}
in the combination of all the images (we have assumed here that the
images $i$ are all independent).  The relevant time ``interval'' is
not the total time spanned by the data but rather
$\delta_t\equiv\sqrt{\var{t}}$, the standard deviation (root variance)
of the times; the best possible proper-motion estimates will have
uncertainties
\begin{equation}
  \sigma_{\mu}\approx\frac{\fwhm}{\delta_t\,\sntotal} \quad ,
  \label{eq:muerror}
\end{equation}
where properly $\fwhm$ is the square-signal-to-noise weighted mean
point-spread function full width at half maximum, and $\delta_t$ is
the square root of the square-signal-to-noise-weighted variance of the
times at which the individual epoch images were taken.

By a similar argument, we hypothesize that the best possible parallax
estimates will have uncertainties
\begin{equation}
  \sigma_{\pi}\approx\frac{\fwhm}{\delta_\lambda\,\sntotal} \quad ,
\end{equation}
where $\delta_\lambda$ is the square root of the
square-signal-to-noise-weighted variance of the trigonometric
functions of the ecliptic longitude $\lambda$ of the Sun (time of year
in angle units):
\begin{equation}
  \delta_\lambda^2\equiv\sigma_{\cos\lambda}^2+\sigma_{\sin\lambda}^2 \quad .
\end{equation}
Essentially, $\delta_\lambda$ describes how well the parallactic
ellipse is sampled; an ideal survey for parallax measurements will
have $\delta_\lambda\approx 1$.

Disk stars move with respect to one another at velocities of $\sim
30\,\kmpers$ \citep{dehnen98a, hogg05a}, that is, on the same order as
the velocity of the Earth around the Sun.  In a multi-epoch survey
spanning a small number of years (such as the SDSS Southern Stripe),
$\delta_t$ is of order unity, so for disk stars the parallax and
proper motion signal-to-noise ratios ought to be comparable in
magnitude.  However, most surveys sample ecliptic longitude $\lambda$
poorly, because of season and scheduling constraints; therefore
$\delta_\lambda$ is usually substantially less than unity, so the
signal-to-noise of parallax is smaller than that of proper motion.

\section{Method}

The goal is to measure the proper motions and parallaxes of sources
detected in multi-epoch data.  We start with a catalog of detections
from a co-addition of the multi-epoch data (co-added at zero lag or
under an assumption that the sources are static).  These detections
serve as ``first-guess'' positions for sources in the imaging.  We
measure the properties of these sources by building models of all the
individual images, at the pixel level, so that each model ``predicts''
every pixel value in every image at every epoch.

Some of the candidate sources will not be point sources but rather
resolved galaxies, and others will not be astronomical sources but
will be caused by artificial satellites or imaging artifacts.  We fit
three qualitatively different models, described below.  One is of a
moving point source, one is of an extended galaxy, and one is of a
general transient or artifact.  For each model, ``fitting''
constitutes optimizing a scalar objective, which is the logarithm of
the likelihood under the assumption that the per-pixel noise is
Gaussian with a known variance in each pixel.  Under the Gaussian
assumption, we can use the different values of the log likelihood to
perform a hypothesis test based on likelihood ratios.  This hypothesis
test distinguishes point sources from extended galaxies and transients
and artifacts.  The parameters of the best-fitting model are the
``measurements'' of the source.

Nothing in what follows fundamentally depends on the assumption of
Gaussian noise.  Data with Poisson errors, for example, can be
analyzed the same way but with the objective function changed to the
logarithm of the Poisson likelihood.  Indeed, any noise model can be
accomodated, though possibly at the expense of computational
simplicity.

In detail, for each source, we have $N$ small images (patches of what
is presumed to be a much larger imaging data set) $i$ taken at times
$t_i$, and we assume that each image has reasonable photometric
calibration, a noise estimate in each pixel (assumed Gaussian, but
that could be relaxed in what follows), and correct astrometric
calibration or world coordinate system (WCS) fixed to an astrometric
$(\RA,\Dec)$ reference frame.  From a co-added image made from all $N$
single-epoch images we have been given a candidate (``first-guess'')
position $(\RA_j,\Dec_j)$ for each source $j$.

\paragraph{point-source model}
The first of the three models is that of a point source, moving in
space and a finite distance from the Solar System.  This point source
is assumed to have a constant flux $S_j$, a position $(\RA_j,\Dec_j)$
at some standard epoch, a parallax $\pi_j$ and a proper motion
$\vecmu_j=(\mualpha_j,\mudelta_j)$.  In this model and the models to
follow, we assume that the sky level has been correctly fitted and
subtracted from the images, or else that sky errors are not strongly
covariant with errors in the model parameters.  In fitting this model,
we find the six-dimensional quantity
$(S_j,\RA_j,\Dec_j,\pi_j,\mualpha_j,\mudelta_j)$ that optimizes the
scalar objective.

Given the times $t_i$ and WCS of the images, any point-source
parameter set \linebreak[4]$(S_j,\RA_j,\Dec_j,\pi_j,\mualpha_j,\mudelta_j)$,
specifies the pixel position of point source $j$ in each image $i$.
This position and the (possibly position-dependent) point-spread
function model for image $i$ permits construction of a pixel-for-pixel
model of source $j$ as it ought to appear in image $i$.

If we had multi-band imaging (the tests below are on are on
single-band images), the flux $S_j$ becomes a set of fluxes $S_{kj}$,
one for each bandpass $k$.  In principle, precise fitting is
complicated by the existence of differential refraction for sources
with extreme colors, so there are relationships among the fluxes
$S_{kj}$, positional offsets, and the airmass or altitude of the
observations.  In the tests below, we are working far enough to the
red that there are no differential refraction issues at the relevant
level of precision.

Although we have assumed non-varying flux in our model, we should
still be able to detect and measure moving sources with varying flux.
We have not investigated this question, but we expect that our method
would produce a flux estimate of approximately the mean flux measured
at the available epochs, and that the point-source model would be
preferred over the transient model, since the objective function is
convex.

\paragraph{galaxy model}
Our model of a resolved galaxy is a Gaussian distribution of flux with
an elliptical covariance parameterized by its radius $r_j$,
eccentricity $e_j$, angle $\theta_j$ and total flux $S_j$.  For each
image, this Gaussian model is convolved with that image's particular
point-spread function to make a seeing-convolved galaxy model.  This
seeing-convolved Gaussian galaxy model is not a realistic galaxy
model, but it is good enough for distinguishing resolved and
unresolved sources at the faint limit, which is sufficient here.
Again, if we had multi-band imaging, the flux $S_j$ would be replaced
by a set of fluxes $S_{jk}$.

\paragraph{junk model}
Our model of a transient or imaging artifact is that there is nothing
but noise in all but one of the images, and that one ``junk'' image
contains many bright pixels.  We compute this model trivially by
computing the chi-squared ($\chi^2$) contribution for each image under
the assumption that there is no flux in the image at all.  The image
with the largest $\chi^2$ contribution is judged to be the ``junk''
image and is discarded.  In order to keep the number of $\chi^2$
contributions constant, we replace the ``junk'' image $\chi^2$ by the
median of the $\chi^2$ contributions of the remaining images.

\paragraph{scalar objective optimization}
The choices of model, scalar objective, and optimization methodology
can all be made independently.  For the objective function the natural
choice is the $\chi^2$ difference between the model and the data taken
over all the pixels that are close to the first-guess position in all
$N$ images.  This objective is analogous to a logarithm of a
likelihood ratio; it is exact if the noise in the image pixels is
Gaussian and independent, with known variances (which can vary from
pixel to pixel).  For optimizing this objective function, we use the
Levenberg-Marquardt method \citep{levenberg44, marquardt63}.

\paragraph{hypothesis test}
In the approximation that the noise is Gaussian, the best fits for
each of the three models can be compared via the best-fit values of
the $\chi^2$ scalar objective.  If the three models are equally likely
\apriori\ and if they have the same number of degrees of freedom, then
one model is confidently preferred over another if it has a best-fit
$\chi^2$ value smaller by an amount $\Delta\chi^2 \gg 1$.  Of course
the models are not equally likely \apriori, but for for the vast
majority of sources, the differences in $\chi^2$ are so large that no
reasonable prior would change the results of our hypothesis test.

Note that there is some degeneracy in our models: a galaxy model with
zero radius and a star model with zero proper motion and parallax
produce exactly the same predictions, and thus our hypothesis test
cannot distinguish them.  This could be remedied by placing prior
probabilies over the model parameters---for example, penalizing tiny
galaxies---but since we are not concerned with the region of parameter
space where this occurs, we have not done this.

Rather than explicitly including a junk model, we could instead place
a threshold on the likelihood of the star and galaxy models: junk data
will be poorly fit by the star and galaxy models and thus will have
tiny likelihood.  In general we have found that image sets for which
the junk model is preferred clearly contain artifacts or transients;
the method is not sensitive to the details of the junk model.

\paragraph{jackknife error analysis}
In principle, the region in parameter space around the best-fit point
where $\chi^2$ is within unity of the minimum provides an estimate of
the uncertainties in the fitted parameters.  However, this estimate is
only good when the model is a good fit; many error contributions in
real data come from source variability, poorly known data properties
(such as pixel uncertainty or point-spread-function estimates that are
in error) and unflagged artifacts in the data.  For this reason, we
use (and advocate) a ``jackknife'' technique for error analysis.

The jackknife technique is to perform the analysis on the $N$ subsets
of the $N$ images created by leaving one image out.  The complete fit
of the three models is performed on each of the $N$ leave-one-out
subsets and parameters are measured.  The uncertainty estimate
$\sigma_p$ for any fitted parameter $p$ is related to the $N$
leave-one-out measurements $p_i$ (made leaving out image $i$) by
\begin{equation}
  \sigma_p^2 = \frac{N-1}{N}\,\sum_i \left(p_i - \mean{p}\right)^2
  \quad ,
  \label{eq:jackknife}
\end{equation}
where $\mean{p}$ is the mean of the leave-one-out measurements $p_i$.
The jackknife technique automatically marginalizes the error estimates
over the other parameters, and provides a properly marginalized
estimate of any multi-parameter covariance matrix by the
generalization of \equationname~(\ref{eq:jackknife}) in which the
square is changed into the $d\times d$ matrix outer product of the
``vectors'' made from the $d$ parameters for which the covariance
matrix is desired.  Of course when $d$ is large, the jackknife will
not accurately sample all degrees of freedom available in the
covariance matrix, but provided $N$ is large enough, it \emph{will}
sample the dominant eigenvectors (the principal components).

\paragraph{Implementation notes}

Our code is implemented in Python and uses the Django web framework,
which provides powerful database and web server integration.  This
allows us to quickly and easily manage and visualize the data and
results.  Combined with the scientific data analysis packages
\code{scipy} and \code{numpy} and the plotting package
\code{matplotlib}, this yielded a powerful software development
environment.

For optimization, we use the Levenberg-Marquardt implementation
\code{levmar} (version 2.2; \citealt{lourakis04}) with Python bindings
\code{pylevmar} (revision 313; \citealt{tse08}).  In this Python
environment, analysis takes on the order of seconds for each source
(30 epochs, $15\times 15$ images), but this could be sped up
substantially by implementing some of the core operations in \code{C}.

\section{Tests on real data}


For test data, we make use of the SDSS Southern Stripe (SDSSSS), a
multi-epoch survey undertaken as part of SDSS-II \citep{adelman08a,
  sdssdr7}.  The SDSSSS data are part of The Sloan Digital Sky Survey
\citep{Gunn:1998vh,York:2000gk}; it involves $ugriz$ CCD imaging of
$\sim 250~\deg^2$ on the Equator in the southern Galactic cap.  All
the SDSSSS data processing, including astrometry \citep{Pier:2002iq},
source identification, deblending and photometry
\citep{Lupton:2001zb}, and calibration \citep{Smith:2002pca,
  padmanabhan08a} are performed with automated SDSS software.

The SDSSSS data have been found to have a small astrometric drift
\citep{bramich08a}, because astrometric calibration was performed at a
single, slightly inappropriate epoch \citep{Pier:2002iq}.  This drift,
for which we are making no correction, is at the $10~\masperyr$ level;
at the precision of this study it does not change any of the
conclusions below.

In general, the hypothesis test we perform requires that the variance
of the noise be properly estimated on a pixel-by-pixel basis.  These
are based on an SDSS imaging noise model, with the adjustment that
pixels that have been corrupted by cosmic rays or other defects are
given infinite variances (vanishing contribution to $\chi^2$).
Occasionally there are unidentified cosmic rays in the data.  These
lead to localized regions with very large contributions to $\chi^2$.
When one of these noise defects appears in the data near one of the
targets, it sometimes causes a source which is truly a galaxy or a
star to be assigned ``junk'' status.  After by-eye inspection of
cutouts, we estimate this rate to be on the order of $<0.5$~percent
for this data source; the rate of such problems increases with the
number of epochs and the image cut-out size (the total number of
pixels in the fit).

For some of the sources we have UKIRT Infrared Deep Sky Survey
(UKIDSS; \citealt{lawrence07a}) data.  UKIDSS uses the UKIRT Wide
Field Camera \citep{casali07a} with the infrared photometric system
described by \cite{hewett06a}, and automated data processing and
archiving \citep{irwin08a, hambly08a}.  The UKIDSS data used here
comes from the fourth data release.

Very red point sources in deep optical imaging---for example,
$z$-band-only sources in the multi-epoch SDSS Southern
Stripe---include both very cool dwarfs and very high redshift quasars.
In principle these can be distinguished with parallax and
proper-motion estimates.  For this reason, we performed a test on
$z$-only point sources in the SDSS Southern Stripe.  The parent sample
is point sources from the SDSSSS ``Co-add Catalog'' (J.~Annis \etal,
in preparation) that have $[i-z]>2~\mag$ and $[r-z]>2~\mag$.  This
criterion selects quasars at $5.8 \la z \la 6.5$ as well as cool
dwarfs with spectral types ranging from mid-L to T (\citealt{fan01a}
and references therein). Hotter brown dwarfs, stars, and
lower-redshift quasars have significant emission in the $i$ band,
giving them bluer $i-z$ colours, while the emission features of cooler
dwarfs and higher-redshift quasars lie mostly redwards of the SDSS $z$
bandpass.

The Co-add Catalog uses the asinh magnitude (or ``luptitude'') scale
\citep{lupton1999}, so it is possible to select objects based on color
even for objects that are not detected in one of the bands.  The
version of the Co-add Catalog we are using is from SDSS Data Release 7
(DR7), and includes $20$ to $40$ epochs over $\sim 250~\deg^2$.

Since we are interested in distinguishing cool brown dwarfs from
high-redshift quasars, we require $\sigma_{\mu}$ to be small
(\equationname\ \ref{eq:muerror}).  We therefore cut our parent sample
to have $z<21~\mag$, leaving roughly 150 sources.  This cut allows us
to reach, with moderate signal-to-noise, slightly fainter sources than
are detectable in the single-epoch images.  In a future paper, we plan
to relax this cut, which should yield considerably more brown dwarf
candidates at smaller signal-to-noise levels.  Of our 150 parent
candidates, some turn out to be caused by an imaging artifact or
transient in one of the $N$ epochs, and some turn out to be galaxies
or stars with mis-measured colors because of data artifacts or
inaccuracies in deblending nearby objects.

Each of the catalog sources has a nominal position and a $z$-band
magnitude in the Co-add Catalog.  For each $z$-only source, we cut out
$15\times 15~\pix^2$ patches of every SDSS image at the nominal
position.  For each tiny image, we construct a tiny local
world-coordinate-system description of the astrometric calibration of
that patch using the SDSS pipeline astrometric calibration.  We
subtract the local value of a smoothly fit sky level (M.~Blanton, in
preparation) and multiply each tiny image by a constant, based on the
pipeline calibration information, to place it on a common photometric
calibration scale in intensity units (energy per unit solid angle per
unit area per unit time per unit frequency).  To the SDSS
pipeline-reconstructed point-spread function (PSF) in each tiny image
we fit a single-Gaussian approximate model, which is not a good fit to
the PSF at high precision, but which is sufficient for modeling
sources at low signal-to-noise.

In this work, we create the $15 \times 15~\pix^2$ cutouts from the
same $20$ to $40$ epochs that are used in the Co-add Catalog; in
future work we plan to use the $\sim 70$ epochs that have become
available in DR7.

We chose $15\times 15~\pix^2$ patches so that a source with proper
motion of $\sim 0.5~\arcsecperyr$ would remain in the patch.  Our
method still works if the source leaves the patch---indeed, our
fastest-moving candidate does this---but we gain no information from
the epochs in which the source has left the patch, so the
signal-to-noise of our parameter estimates will be less than optimal
when this happens.  We could choose to use larger cutouts; the only
difficulty is that if the patch contains more than one source, our
model will try to explain the brightest source (because this will
decrease $\chi^2$ the most).  This could perhaps be remedied by adding
a prior on the source position, but since the SDSSSS data are from
well below the galactic equator, stellar density is low and we have
not found this to be necessary.  Alternatively, in cases where the
source leaves the original patch we could produce new cutouts that
track the source motion.

\figurename s~\ref{fig:example} through \ref{fig:examplecrap}
illustrate our approach by showing the results of the (moving)
point-source and (static) galaxy model fits to four sources in the the
SDSSSS data.  In these figures we show all the individual $15\times
15$ images from the individual epochs, and the best-fit point-source
and galaxy parameters.  In these figures, we visualize the
distribution of acceptable parameters around the best-fit values
through sampling.  We also show mean images and mean residual maps in
the static and moving coordinate systems.  These figures demonstrate
heuristically that the hypothesis test is effective at separating
sources of different types, even when the source is not apparent at
high signal-to-noise at any individual epoch.

In \figurename~\ref{fig:bubbles} we show the overall results from
application of our techniques to the $[i-z]>2$~mag sources in the
SDSSSS: We show proper-motion measurements and jackknife estimates of
our uncertainties as a function of $z$-band magnitude.  Known quasars
and brown dwarfs are marked.  Our measurements clearly separate the
known quasars and brown dwarfs on the basis of proper motion alone.
All known brown dwarfs in the sample obtain significant non-zero
proper motion measurements, and all known high-redshift quasars in our
sample obtain proper motion measurements consistent with zero.  The
sources in our sample that have significant motions and have not been
previously identified as brown dwarfs are our new brown dwarf
candidates.  In \figurename~\ref{fig:colormag} we show the UKIDSS and
SDSS $[z-J]$ colors of the sources for which we have UKIDSS $J$
measurements, with the known brown dwarfs and quasars and our new
brown dwarf candidates marked.  
Many of the sources in these \figurename s are undetectable (or not
detectable reliably) at individual epochs; the single-epoch $5$-sigma
detection limit is roughly $z=20.5~\mag$ in good seeing conditions
\citep{sdssdr7}.

In \figurename~\ref{fig:info}, our jackknife estimates of our
measurement uncertainties are compared to approximate estimates of the
total information content in each source's data set, made with an
approximation to \equationname~(\ref{eq:muerror}).  If our uncertainty
estimates are correct (as we demonstrate that they are, below), this
shows that we come close to attaining the accuracy available.

\section{Tests on artificial data}

To demonstrate that our jackknife error estimates are reasonable, and
that our code is optimizing the models correctly, we performed some
tests on synthetic data.  We selected a subset of the SDSSSS candidate
objects for which we found reasonable fits to a moving point source
model.  For each candidate, we generated a stack of images by
generating, for each image in the original stack, the image predicted
by our point-source model, given the WCS, point-spread function, time,
and noise amplitude of the image.  This is a good test set because it
has the same imaging properties as the original data and the same
distribution of point-source parameters as the sources we want to be
able to discover.  Since the synthetic images are generated using our
image model, this test shows how our algorithm would perform if our
modeling assumptions were exactly correct.

After running our optimization code on these synthetic images, we
compare our errors---the differences between the true and estimated
moving-point-source parameters---to the jackknife estimates of our
uncertainties.  In \figurename~\ref{fig:fake} we show that the errors
are consistent with the uncertainty estimates.  This shows that when
our assumptions about the data are correct, we do measure the proper
motions as accurately as our jackknife errors indicate.

\section{Discussion}

We have shown that straightforward image modeling permits the
measurement of apparent motions, especially the proper motion and
parallax of a source in multi-epoch data, even when the source is too
faint to be reliably detected or centroided at any individual epoch.
The results of this project are not surprising; indeed what is
surprising is how rarely the measurements of stellar motions are made
by comprehensive data modeling.

We demonstrated the technique on real and artificial data.  In the
process of performing these tests we showed that spectrosopically
confirmed quasars and brown dwarfs can be perfectly distinguished with
proper motions measured by this technique.  Working without proper
motions, but with Co-add Catalog sources and a significant amount of
near-infrared imaging follow-up, a group has followed up the $z$-only
sources most likely to be high-redshift quasars \citep{chiu08a,
jiang08a}.  This project, even after infrared imaging, found---after
expensive spectroscopic follow-up---that some of the high-redshift
quasar candidates selected on the basis of visible and near-infrared
imaging are in fact nearby brown dwarfs.  We have shown that all of
these spectroscopically confirmed brown dwarfs have significantly
measured ($>5$~sigma) non-zero proper motions by the technique shown
here (and are reported in \tablename~\ref{tab:movingsources}).  None
of the spectroscopically confirmed high-redshift quasars do.  Use of
this technique could have been used to substantially increase the
efficiency of either quasar or brown-dwarf searches in this data set.

In performing this demonstration, we have independently identified
all 10 known brown dwarfs \citep{fan00a, geballe02a, hawley02a,
berriman03a, knapp04a, chiu08a, metchev08a} in our parent sample, and
we have discovered 9 \emph{new} candidate brown dwarfs, presented in
\tablename~\ref{tab:movingsources}.  Based on our analysis, these
objects have a high probability of being brown dwarfs.  It would be
desirable to separate disk dwarfs from halo dwarfs---the fastest
angular movers tend to be halo members (for example,
\citealt{lepine03a})---but the time cadence of the SDSSSS data is such
that parallaxes are not measured well.  Two of the dwarfs we
rediscover---2MASS J010752.42+004156.3 and 2MASS
J020742.84+000056.4---have previously measured parallaxes
\citep{vrba04a}; the measurements are consistent with our upper
limits.

Our tests show that the uncertainty in the proper-motion measurement
made by image modeling is consistent with the best possible
uncertainties given the angular resolution and photometric sensitivity
of the combination of all images in the multi-epoch data set.  These
tests effectively show that such measurements can be made for objects
that are fainter than those available to traditional methods that
require source detection at every epoch.  In imaging with $N$ equally
sensitive epochs, we are able to measure objects that are fainter by
$\Delta m$ magnitudes:
\begin{eqnarray}
 \Delta m &=& -\log_{2.5}(\sn_i) + \log_{2.5}(\sntotal) \\
 &=& \log_{2.5}(\sqrt{N}) \\
 &\sim& 0.55\,\log N~\mag \quad .
\end{eqnarray}
This advantage amounts to $1~\mag$ for surveys with $6$ similar
epochs, and $1.6$ to $2.0~\mag$ in data with $20$ to $40$ epochs (such
as the data used here).  In the $\sim 70$ epochs available in SDSS
DR7, it reaches $2.3~\mag$.  Several of the high-redshift quasars and
brown dwarfs analyzed in this study were only detectable in the
combination of all of the multi-epoch images.

The depth advantage of image modeling is most dramatic in surveys with
very large numbers of epochs, as is expected for LSST.  In general the
number of interesting sources is a strong function of depth (factors
of $2$ to $4$ per magnitude), so the ``reach'' of the image-modeling
technique is a strong function of the number of epochs.

One limitation of the work presented here is that we used the
zero-proper-motion image ``stack'' for source detection and therefore
will only have in the candidate list objects with small proper
motions.  Faint stars and dwarfs with proper motions large enough that
they move the width of the PSF between epochs, or some significant
fraction of that, are harder to find, because they don't appear in the
stack at much higher signal-to-noise than they appear in any
individual-epoch image.  In future work we hope to address the
detection and measurement of these fast-moving but very faint sources.
Approximations have been executed in the search for Solar System
bodies (for example, \citealt{bernstein04a}).  Certainly a reliable
system for discovery in this regime would have a big impact on future
surveys like PanSTARRS and LSST.

\acknowledgements We thank Jon Barron, Mike Blanton, Bertrand Goldman,
Linhua Jiang, Keir Mierle, Sam Roweis, Iain Murray, Ralf-Dieter
Scholz, Christopher Kochanek, and Rob Fergus for help, comments, and
software.  We thank our anonymous reviewer for detailed and thoughtful
comments which greatly improved this paper.  This project was
partially supported by the US National Science Foundation (grant
AST-0428465) and the US National Aeronautics and Space Administration
(grants NAG5-11669 and 07-ADP07-0099).  During part of the period in
which this research was performed, DWH was a research fellow of the
German Alexander von Humboldt Foundation.

This project made use of the SDSS Southern Stripe Co-add Catalog, which
was constructed by Jim Annis, Huan Lin, Robert Lupton, and others, who
graciously made it available to us in advance of publication.

Funding for the SDSS and SDSS-II has been provided by the Alfred
P. Sloan Foundation, the Participating Institutions, the National
Science Foundation, the U.S. Department of Energy, the National
Aeronautics and Space Administration, the Japanese Monbukagakusho, the
Max Planck Society, and the Higher Education Funding Council for
England. The SDSS Web Site is http://www.sdss.org/.

The SDSS is managed by the Astrophysical Research Consortium for the
Participating Institutions. The Participating Institutions are the
American Museum of Natural History, Astrophysical Institute Potsdam,
University of Basel, University of Cambridge, Case Western Reserve
University, University of Chicago, Drexel University, Fermilab, the
Institute for Advanced Study, the Japan Participation Group, Johns
Hopkins University, the Joint Institute for Nuclear Astrophysics, the
Kavli Institute for Particle Astrophysics and Cosmology, the Korean
Scientist Group, the Chinese Academy of Sciences, Los Alamos National
Laboratory, the Max-Planck-Institute for Astronomy, the
Max-Planck-Institute for Astrophysics, New Mexico State University,
Ohio State University, University of Pittsburgh, University of
Portsmouth, Princeton University, the United States Naval Observatory,
and the University of Washington.

This work is based in part on data obtained as part of the UKIRT
Infrared Deep Sky Survey.  The United Kingdom Infrared Telescope is
operated by the Joint Astronomy Centre on behalf of the Science and
Technology Facilities Council of the U.K.

This publication makes use of data products from the Two Micron All
Sky Survey, which is a joint project of the University of
Massachusetts and the Infrared Processing and Analysis
Center/California Institute of Technology, funded by the National
Aeronautics and Space Administration and the National Science
Foundation.

This research made use of the NASA Astrophysics Data System and the
WFCAM Science Archive.  This research has benefitted from the M, L,
and T dwarf compendium housed at \linebreak[4]
\code{DwarfArchives.org} and maintained by Chris Gelino, Davy
Kirkpatrick, and Adam Burgasser.  This research made use of the
\code{idlutils} and \code{photoop} software suites (maintained by
David Schlegel, Nikhil Padmanabhan, Doug Finkbeiner, Mike Blanton, and
others) and the Python programming language and Python packages
\code{scipy}, \code{matplotlib} and \code{Django}.

\clearpage
\begin{table}
\tiny\begin{tabular}{rrrrrrrl}
\multicolumn{1}{c}{Name} & \multicolumn{1}{c}{RA} & \multicolumn{1}{c}{Dec} & \multicolumn{1}{c}{flux} & \multicolumn{1}{c}{parallax} & \multicolumn{1}{c}{dRA/d$t$} & \multicolumn{1}{c}{dDec/d$t$} & \multicolumn{1}{c}{notes} \\
\multicolumn{1}{c}{} & \multicolumn{1}{c}{deg} & \multicolumn{1}{c}{deg} & \multicolumn{1}{c}{$s/n$} & \multicolumn{1}{c}{arcsec} & \multicolumn{1}{c}{arcsec\,yr$^{-1}$} & \multicolumn{1}{c}{arcsec\,yr$^{-1}$} & \multicolumn{1}{c}{} \\ \hline
SDSS~J001608.47$-$004302.9 & $  4.03537$ & $-0.71733$ & $ 46.4$ & $+0.007\pm0.058$ & $+0.134\pm0.015$ & $-0.020\pm0.007$ & 2 U BD \\
SDSS~J005212.29+001216.0 & $ 13.05131$ & $+0.20465$ & $ 29.6$ & $-0.053\pm0.069$ & $-0.165\pm0.014$ & $-0.211\pm0.010$ & 2 U BD \\
SDSS~J010407.68$-$005329.1 & $ 16.03195$ & $-0.89128$ & $ 59.1$ & $-0.041\pm0.068$ & $+0.460\pm0.010$ & $-0.017\pm0.008$ & 2 U BD \\
SDSS~J010752.59+004156.0 & $ 16.96899$ & $+0.69905$ & $ 59.0$ & $+0.044\pm0.070$ & $+0.644\pm0.014$ & $+0.085\pm0.013$ & 2 \makebox[\widthof{U}]{} BD \\
SDSS~J020333.28$-$010813.1 & $ 30.88881$ & $-1.13683$ & $ 18.6$ & $-0.019\pm0.063$ & $+0.354\pm0.019$ & $-0.005\pm0.012$ & \makebox[\widthof{2}]{} U BD \\
SDSS~J020742.85+000055.6 & $ 31.92867$ & $+0.01561$ & $ 27.8$ & $+0.078\pm0.072$ & $+0.163\pm0.017$ & $-0.029\pm0.010$ & 2 U BD \\
SDSS~J023617.95+004853.5 & $ 39.07492$ & $+0.81501$ & $ 74.4$ & $+0.025\pm0.021$ & $+0.134\pm0.006$ & $-0.166\pm0.005$ & 2 U BD \\
SDSS~J033035.23$-$002537.2 & $ 52.64687$ & $-0.42678$ & $ 76.5$ & $+0.040\pm0.026$ & $+0.390\pm0.007$ & $-0.360\pm0.008$ & 2 U BD \\
SDSS~J214046.48+011258.2 & $325.19385$ & $+1.21633$ & $ 28.3$ & $-0.064\pm0.143$ & $-0.085\pm0.012$ & $-0.215\pm0.007$ & 2 \makebox[\widthof{U}]{} BD \\
SDSS~J224953.45+004403.9 & $342.47285$ & $+0.73458$ & $ 44.4$ & $-0.003\pm0.142$ & $+0.084\pm0.010$ & $+0.011\pm0.009$ & 2 U BD \\

\hline
SDSS~J001836.46$-$002559.9 & $  4.65204$ & $-0.43315$ & $ 27.6$ & $+0.038\pm0.137$ & $+0.179\pm0.019$ & $-0.029\pm0.017$ & \makebox[\widthof{2}]{} U  \\
SDSS~J011014.40+010618.5 & $ 17.55990$ & $+1.10534$ & $ 32.0$ & $-0.017\pm0.089$ & $+0.542\pm0.023$ & $+0.013\pm0.015$ & \makebox[\widthof{2}]{} U  \\
SDSS~J011417.92$-$003437.9 & $ 18.57481$ & $-0.57704$ & $ 19.4$ & $+0.090\pm0.064$ & $-0.093\pm0.021$ & $-0.077\pm0.018$ & \makebox[\widthof{2}]{} U  \\
SDSS~J021642.94+004005.1 & $ 34.17907$ & $+0.66828$ & $ 52.0$ & $-0.021\pm0.023$ & $-0.069\pm0.011$ & $-0.093\pm0.009$ & \makebox[\widthof{2}]{} U  \\
SDSS~J023047.97$-$002600.4 & $ 37.69996$ & $-0.43332$ & $ 29.9$ & $-0.012\pm0.045$ & $+0.127\pm0.008$ & $-0.003\pm0.010$ & \makebox[\widthof{2}]{} U  \\
SDSS~J215919.95+003309.0 & $329.83326$ & $+0.55260$ & $ 21.5$ & $+0.120\pm0.158$ & $+0.155\pm0.025$ & $+0.100\pm0.018$ & \makebox[\widthof{2}]{} \makebox[\widthof{U}]{}  \\
SDSS~J234730.64$-$002912.0 & $356.87782$ & $-0.48653$ & $ 14.7$ & $+0.095\pm0.120$ & $-0.082\pm0.020$ & $-0.090\pm0.026$ & \makebox[\widthof{2}]{} U  \\
SDSS~J234841.38$-$004022.9 & $357.17250$ & $-0.67289$ & $ 39.7$ & $+0.161\pm0.084$ & $+0.097\pm0.025$ & $-0.125\pm0.035$ & \makebox[\widthof{2}]{} U  \\
SDSS~J235410.42+004315.9 & $358.54362$ & $+0.72131$ & $ 59.6$ & $-0.061\pm0.077$ & $+0.053\pm0.013$ & $-0.063\pm0.010$ & \makebox[\widthof{2}]{} U  \\
\end{tabular}
\caption{Well-fit $[i-z]>2$~mag sources in the SDSS Southern Stripe
  with proper motions \mbox{$>60~\masperyr$} measured at high
  confidence ($>3.5$~sigma).  RA, Dec positions have equinox J2000.0
  but are computed for MJD 53000.  The note ``2'' indicates that there
  is a nearby entry in the 2MASS point-source catalog
  \citep{skrutskie06a}, while ``U'' indicates a nearby source in the
  UKIDSS catalog \citep{lawrence07a}.  ``BD'' indicates objects that
  are spectroscopically-confirmed brown dwarfs \citep{fan00a,
  geballe02a, hawley02a, berriman03a, knapp04a, chiu08a, metchev08a}.
  The 9 sources in the lower part of the table (not marked with
  ``BD'') are new brown-dwarf candidates.\label{tab:movingsources}}
\end{table}

\clearpage
\begin{table}
\tiny\begin{tabular}{rrrrrr}
\multicolumn{1}{c}{Name} & \multicolumn{1}{c}{SDSS u} & \multicolumn{1}{c}{SDSS g} & \multicolumn{1}{c}{SDSS r} & \multicolumn{1}{c}{SDSS i} & \multicolumn{1}{c}{SDSS z} \\
\multicolumn{1}{c}{} & \multicolumn{1}{c}{mag} & \multicolumn{1}{c}{mag} & \multicolumn{1}{c}{mag} & \multicolumn{1}{c}{mag} & \multicolumn{1}{c}{mag} \\ \hline
SDSS~J001608.47$-$004302.9 & $ 24.26 \pm   0.29$ & $ 25.13 \pm   0.16$ & $ 23.88 \pm   0.18$ & $ 21.31 \pm   0.03$ & $ 19.23 \pm   0.01$ \\
SDSS~J005212.29+001216.0 & $ 24.35 \pm   0.26$ & $ 25.03 \pm   0.13$ & $ 23.88 \pm   0.10$ & $ 21.92 \pm   0.03$ & $ 19.62 \pm   0.02$ \\
SDSS~J010407.68$-$005329.1 & $ 25.02 \pm   0.23$ & $ 25.17 \pm   0.22$ & $ 24.12 \pm   0.13$ & $ 22.07 \pm   0.04$ & $ 19.83 \pm   0.02$ \\
SDSS~J010752.59+004156.0 & $ 24.31 \pm   0.21$ & $ 25.17 \pm   0.13$ & $ 24.09 \pm   0.11$ & $ 21.87 \pm   0.03$ & $ 19.11 \pm   0.01$ \\
SDSS~J020333.28$-$010813.1 & $ 24.60 \pm   0.24$ & $ 25.18 \pm   0.21$ & $ 24.60 \pm   0.14$ & $ 23.91 \pm   0.14$ & $ 20.85 \pm   0.04$ \\
SDSS~J020742.85+000055.6 & $ 24.43 \pm   0.25$ & $ 25.25 \pm   0.13$ & $ 25.04 \pm   0.13$ & $ 24.15 \pm   0.13$ & $ 20.43 \pm   0.04$ \\
SDSS~J023617.95+004853.5 & $ 24.83 \pm   0.20$ & $ 24.89 \pm   0.12$ & $ 23.85 \pm   0.09$ & $ 21.69 \pm   0.03$ & $ 19.01 \pm   0.01$ \\
SDSS~J033035.23$-$002537.2 & $ 24.66 \pm   0.25$ & $ 25.01 \pm   0.13$ & $ 23.06 \pm   0.05$ & $ 20.88 \pm   0.01$ & $ 18.79 \pm   0.01$ \\
SDSS~J214046.48+011258.2 & $ 24.60 \pm   0.29$ & $ 24.67 \pm   0.15$ & $ 23.55 \pm   0.11$ & $ 21.16 \pm   0.02$ & $ 19.11 \pm   0.01$ \\
SDSS~J224953.45+004403.9 & $ 24.01 \pm   0.21$ & $ 25.24 \pm   0.15$ & $ 23.79 \pm   0.11$ & $ 21.61 \pm   0.03$ & $ 19.53 \pm   0.01$ \\

\hline
SDSS~J001836.46$-$002559.9 & $ 24.65 \pm   0.27$ & $ 25.09 \pm   0.14$ & $ 24.29 \pm   0.19$ & $ 23.23 \pm   0.11$ & $ 20.43 \pm   0.03$ \\
SDSS~J011014.40+010618.5 & $ 24.95 \pm   0.25$ & $ 25.27 \pm   0.13$ & $ 24.28 \pm   0.13$ & $ 22.29 \pm   0.05$ & $ 20.03 \pm   0.02$ \\
SDSS~J011417.92$-$003437.9 & $ 24.55 \pm   0.32$ & $ 25.10 \pm   0.13$ & $ 24.76 \pm   0.15$ & $ 22.46 \pm   0.09$ & $ 20.40 \pm   0.03$ \\
SDSS~J021642.94+004005.1 & $ 24.82 \pm   0.20$ & $ 24.98 \pm   0.12$ & $ 24.14 \pm   0.11$ & $ 22.16 \pm   0.04$ & $ 20.01 \pm   0.02$ \\
SDSS~J023047.97$-$002600.4 & $ 24.52 \pm   0.25$ & $ 24.89 \pm   0.12$ & $ 24.13 \pm   0.11$ & $ 22.36 \pm   0.04$ & $ 20.22 \pm   0.02$ \\
SDSS~J215919.95+003309.0 & $ 24.72 \pm   0.25$ & $ 24.89 \pm   0.15$ & $ 24.20 \pm   0.15$ & $ 22.59 \pm   0.09$ & $ 20.58 \pm   0.04$ \\
SDSS~J234730.64$-$002912.0 & $ 24.46 \pm   0.28$ & $ 25.03 \pm   0.14$ & $ 24.78 \pm   0.15$ & $ 22.93 \pm   0.08$ & $ 20.90 \pm   0.05$ \\
SDSS~J234841.38$-$004022.9 & $ 24.76 \pm   0.29$ & $ 25.07 \pm   0.16$ & $ 23.93 \pm   0.12$ & $ 21.95 \pm   0.04$ & $ 19.93 \pm   0.02$ \\
SDSS~J235410.42+004315.9 & $ 24.04 \pm   0.19$ & $ 25.03 \pm   0.22$ & $ 23.24 \pm   0.10$ & $ 21.17 \pm   0.07$ & $ 19.11 \pm   0.01$ \\
\end{tabular}
\vspace{12pt}

\tiny\begin{tabular}{rrrrrrrr}
\multicolumn{1}{c}{Name} & \multicolumn{1}{c}{UKIDSS y} & \multicolumn{1}{c}{2MASS J} & \multicolumn{1}{c}{UKIDSS J1} & \multicolumn{1}{c}{2MASS H} & \multicolumn{1}{c}{UKIDSS H} & \multicolumn{1}{c}{2MASS K$_{s}$} & \multicolumn{1}{c}{UKIDSS K} \\
\multicolumn{1}{c}{} & \multicolumn{1}{c}{mag} & \multicolumn{1}{c}{mag} & \multicolumn{1}{c}{mag} & \multicolumn{1}{c}{mag} & \multicolumn{1}{c}{mag} & \multicolumn{1}{c}{mag} & \multicolumn{1}{c}{mag} \\ \hline
SDSS~J001608.47$-$004302.9 & $ 17.69 \pm   0.02$ & $ 16.33 \pm   0.12$ & $ 16.30 \pm   0.01$ & $ 15.23 \pm   0.11$ & $ 15.35 \pm   0.01$ & $ 14.54 \pm   0.09$ & $ 14.50 \pm   0.01$ \\
SDSS~J005212.29+001216.0 & $ 17.87 \pm   0.03$ & $ 16.36 \pm   0.11$ & $ 16.54 \pm   0.02$ & $ 15.56 \pm   0.13$ & $ 15.81 \pm   0.01$ & $ 15.46 \pm   0.16$ & $ 15.21 \pm   0.01$ \\
SDSS~J010407.68$-$005329.1 & $ 17.87 \pm   0.03$ & $ 16.53 \pm   0.13$ & $ 16.54 \pm   0.01$ & $ 15.64 \pm   0.14$ & $ 15.80 \pm   0.02$ & $ 15.33 \pm   0.17$ & $ 15.24 \pm   0.02$ \\
SDSS~J010752.59+004156.0 &  & $ 15.82 \pm   0.06$ &  & $ 14.51 \pm   0.04$ &  & $ 13.71 \pm   0.04$ &  \\
SDSS~J020333.28$-$010813.1 & $ 18.99 \pm   0.08$ &  & $ 17.70 \pm   0.04$ &  & $ 16.89 \pm   0.03$ &  & $ 16.26 \pm   0.03$ \\
SDSS~J020742.85+000055.6 & $ 18.03 \pm   0.03$ & $ 16.80 \pm   0.16$ & $ 16.74 \pm   0.01$ & $ 16.40$ & $ 16.81 \pm   0.04$ & $ 15.41$ & $ 16.73 \pm   0.05$ \\
SDSS~J023617.95+004853.5 & $ 17.18 \pm   0.02$ & $ 16.10 \pm   0.08$ &  & $ 15.27 \pm   0.07$ & $ 15.14 \pm   0.01$ & $ 14.67 \pm   0.09$ & $ 14.56 \pm   0.01$ \\
SDSS~J033035.23$-$002537.2 & $ 16.50 \pm   0.01$ & $ 15.31 \pm   0.05$ & $ 15.21 \pm   0.00$ & $ 14.42 \pm   0.04$ & $ 14.49 \pm   0.00$ & $ 13.84 \pm   0.05$ & $ 13.77 \pm   0.00$ \\
SDSS~J214046.48+011258.2 &  & $ 15.89 \pm   0.08$ &  & $ 15.31 \pm   0.09$ &  & $ 14.42 \pm   0.08$ &  \\
SDSS~J224953.45+004403.9 & $ 17.78 \pm   0.03$ & $ 16.59 \pm   0.12$ & $ 16.45 \pm   0.01$ & $ 15.42 \pm   0.11$ & $ 15.33 \pm   0.01$ & $ 14.36 \pm   0.07$ & $ 14.41 \pm   0.01$ \\

\hline
SDSS~J001836.46$-$002559.9 & $ 18.73 \pm   0.08$ &  & $ 17.67 \pm   0.05$ &  & $ 16.61 \pm   0.04$ &  &  \\
SDSS~J011014.40+010618.5 &  &  & $ 16.90 \pm   0.02$ &  & $ 16.34 \pm   0.03$ &  & $ 15.73 \pm   0.02$ \\
SDSS~J011417.92$-$003437.9 & $ 18.85 \pm   0.08$ &  & $ 17.89 \pm   0.04$ &  & $ 17.43 \pm   0.07$ &  & $ 16.73 \pm   0.06$ \\
SDSS~J021642.94+004005.1 & $ 18.41 \pm   0.05$ &  & $ 17.31 \pm   0.03$ &  & $ 16.97 \pm   0.04$ &  & $ 16.51 \pm   0.04$ \\
SDSS~J023047.97$-$002600.4 &  &  & $ 17.35 \pm   0.03$ &  & $ 16.64 \pm   0.04$ &  & $ 16.08 \pm   0.04$ \\
SDSS~J215919.95+003309.0 &  &  &  &  &  &  &  \\
SDSS~J234730.64$-$002912.0 & $ 19.56 \pm   0.11$ &  & $ 18.37 \pm   0.08$ &  & $ 17.87 \pm   0.09$ &  & $ 17.18 \pm   0.08$ \\
SDSS~J234841.38$-$004022.9 &  &  & $ 17.50 \pm   0.03$ &  & $ 16.89 \pm   0.03$ &  & $ 16.40 \pm   0.04$ \\
SDSS~J235410.42+004315.9 & $ 17.82 \pm   0.03$ &  & $ 17.01 \pm   0.02$ &  & $ 16.38 \pm   0.03$ &  & $ 15.96 \pm   0.03$ \\
\end{tabular}
\caption[Photometric properties for the fast-moving objects.]
{Photometric properties for the sources in
\tablename~\ref{tab:movingsources}, by association with sources in
SDSS, 2MASS, and UKIDSS.  SDSS magnitudes are given in the SDSS
system, which is very close to the AB system (See
\url{http://www.sdss.org/dr7/algorithms/fluxcal.html\#sdss2ab});
UKIDSS magnitudes are given as retrieved from the WFCAM science
archive, \ie, they are Vega-based.\label{tab:photometry}}
\end{table}


\clearpage
\begin{figure}
\resizebox{\examplefigurewidth}{!}{\includegraphics{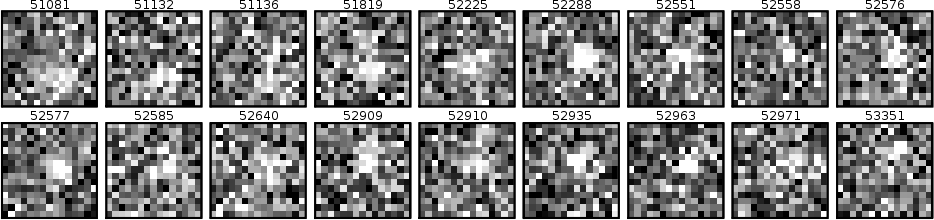}}\\[2ex]%
\resizebox{\examplefigurewidth}{!}{\includegraphics{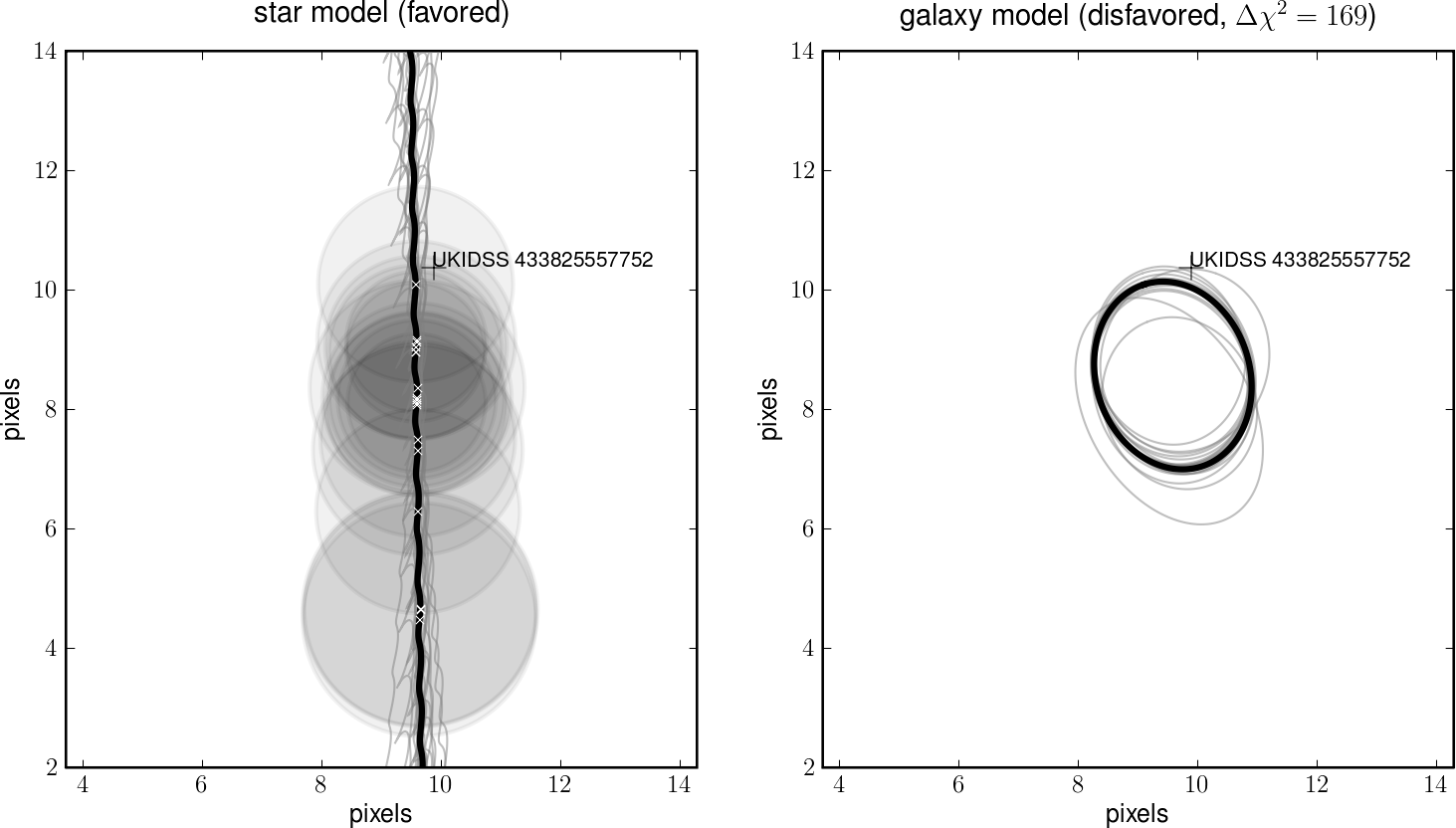}}\\[2ex]%
\resizebox{\examplefigurewidth}{!}{\includegraphics{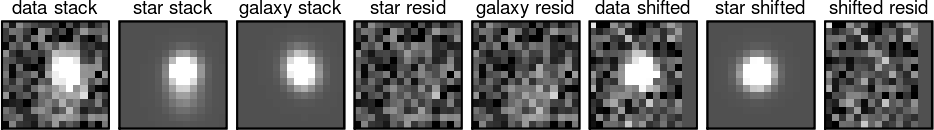}}
\caption{[caption on next page]}
\end{figure}

\clearpage
\addtocounter{figure}{-1}
\begin{figure}
  \caption{[on previous page] The results of fitting the SDSS Southern
    Stripe multi-epoch imaging data on SDSS J020333.28-010813.1, a
    spectroscopically confirmed brown dwarf (\citealt{knapp04a},
    L.~Jiang, private communication) and $[i-z]>2$~mag source.  The
    top set of panels---labeled by observation MJD---show the
    individual epoch $15\times 15$-pixel sub-images; note that the
    source is not clearly detectable at every epoch.  The middle
    diagrams show the output of fitting a moving source (left panel)
    or a resolved galaxy (right panel).  On the moving-source diagram,
    the best fit path of the moving point source is shown as a thick
    black line, the thinner grey lines show alternative paths sampled
    from the jackknife-inferred posterior distribution of trajectories
    consistent with the data; that is, the grey lines effectively show
    the uncertainty interval.  The thick black and thin grey lines
    contain wiggles with a period of one year (or the pixel distance
    of one year at that path's proper motion) because each is the
    realization of a trajectory with finite proper motion and
    parallax.  It can be seen from this panel that this source has a
    well-measured proper motion but not a well-measured parallax,
    because the grey lines do not share a common parallax.  The PSF
    FWHM sizes of the individual epoch images are shown as circles
    centered on the positions the point source would have on the
    best-fit path.  On the galaxy diagram, the mean-PSF-convolved
    galaxy model is shown as a black ellipse, and the grey ellipses
    sample the jackknife-inferred distribution of galaxy models
    consistent with the data.  Note that for this source, the
    point-source model is a much better fit than the galaxy model
    (with $\chi^2$ difference $169$), so the point-source model is
    favored.  Along the bottom, the leftmost panel (data stack) shows
    the data co-added (weighted by per-pixel inverse variance).  The
    second and third panels (star stack and galaxy stack) show the
    star and galaxy models co-added at zero lag (no proper motion
    compensation).  The fourth panel (star resid) shows the residuals
    (data minus model) for the point-source model, and the fifth panel
    (galaxy resid) the residuals for the galaxy model, both co-added
    at zero lag.  The sixth panel (data shifted) shows the data
    co-added with the best-fit proper motion compensated.  The seventh
    (star shifted) shows the point-source model, co-added with the
    best-fit proper motion compensated.  The final panel (shifted
    resid) shows the residuals for the point-source model co-added
    with the best-fit proper motion compensated.  The
    motion-compensating shifts are rounded to the nearest pixel.  Note
    that these co-added figures are only shown for purposes of
    illustration; our method \emph{never} co-adds
    images.\label{fig:example}}
\end{figure}

\clearpage
\begin{figure}
\resizebox{\examplefigurewidth}{!}{\includegraphics{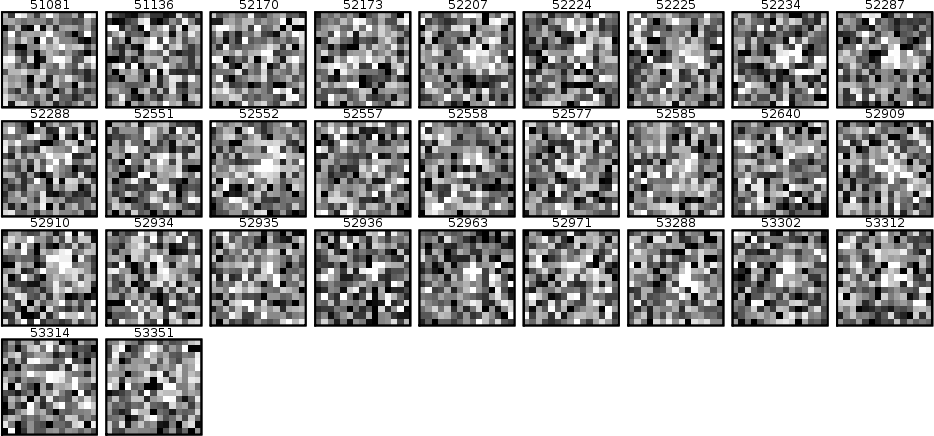}}\\[2ex]%
\resizebox{\examplefigurewidth}{!}{\includegraphics{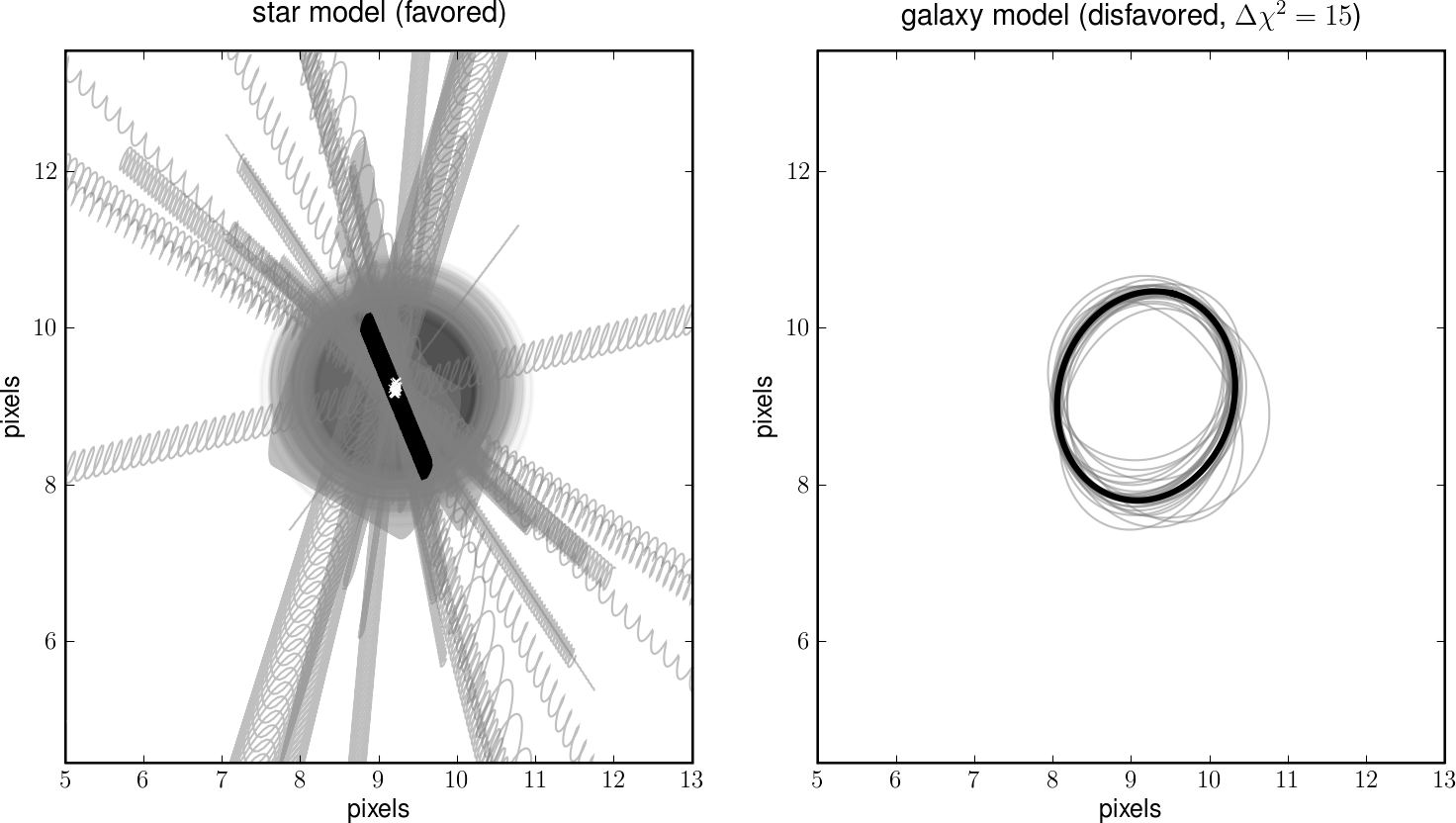}}\\[2ex]%
\resizebox{\examplefigurewidth}{!}{\includegraphics{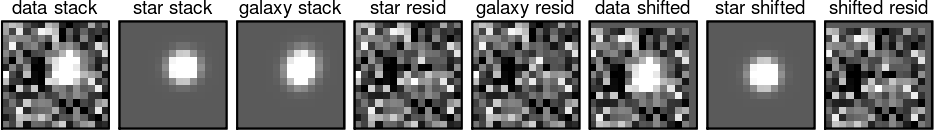}}
\caption{The same as \figurename~\ref{fig:example} but for
  SDSS~J020332.35+001228.6, a spectroscopically confirmed $z\sim 6$
  quasar \citep{jiang08a} and $[i-z]>2$~mag source.  Here the
  point-source model is favored, but the inferred proper motion
  (best-fit value or any sample from the distribution) is very small;
  the wavy paths each span $\sim 100~\yr$ in time; they have different
  position angles because when the magnitude of the proper motion is
  constrained to be near zero, the direction is not well constrained.
  The galaxy model is disfavored by a small amount.  The amount is
  small because the best-fit galaxy model is a non-moving compact
  source, which is not dissimilar to the nearly non-moving best-fit
  point source.\label{fig:exampleQSO}}
\end{figure}

\clearpage
\begin{figure}
\resizebox{\examplefigurewidth}{!}{\includegraphics{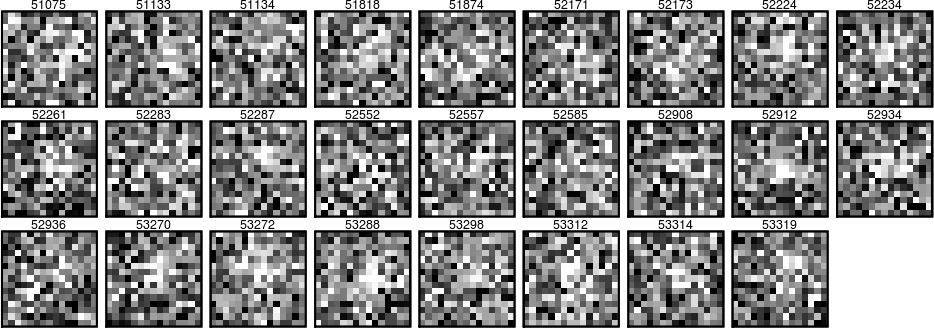}}\\[2ex]%
\resizebox{\examplefigurewidth}{!}{\includegraphics{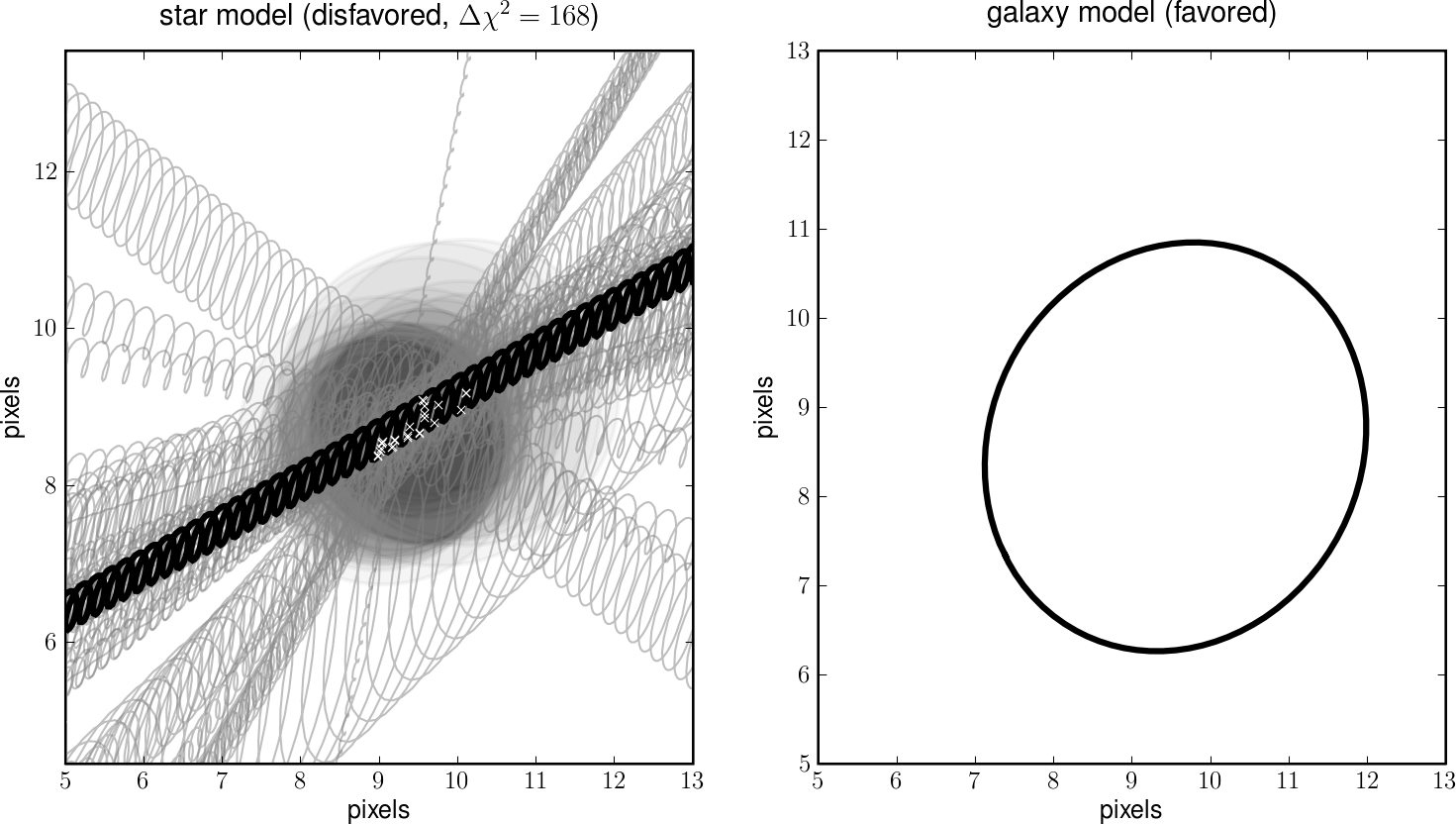}}\\[2ex]%
\resizebox{\examplefigurewidth}{!}{\includegraphics{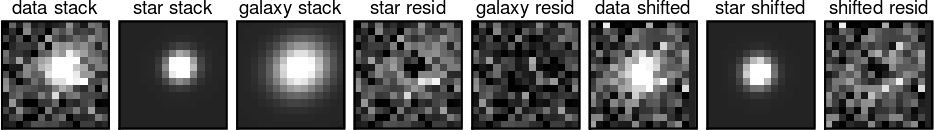}}
\caption{The same as \figurename~\ref{fig:example} but for a faint
  galaxy.  Here the galaxy model is favored.\label{fig:examplegalaxy}}
\end{figure}

\clearpage
\begin{figure}
\resizebox{\examplefigurewidth}{!}{\includegraphics{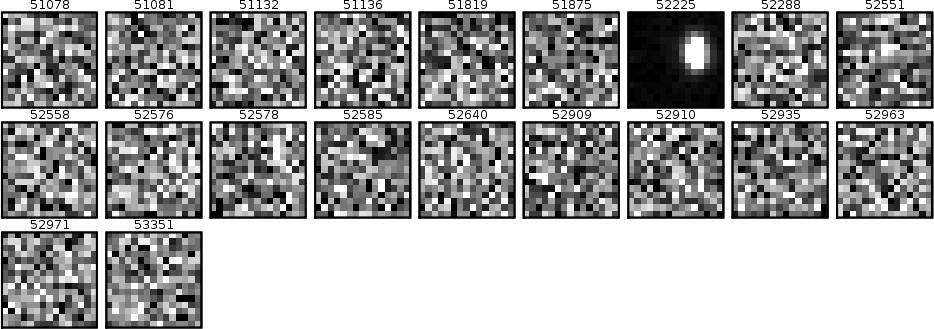}}\\[2ex]%
\resizebox{\examplefigurewidth}{!}{\includegraphics{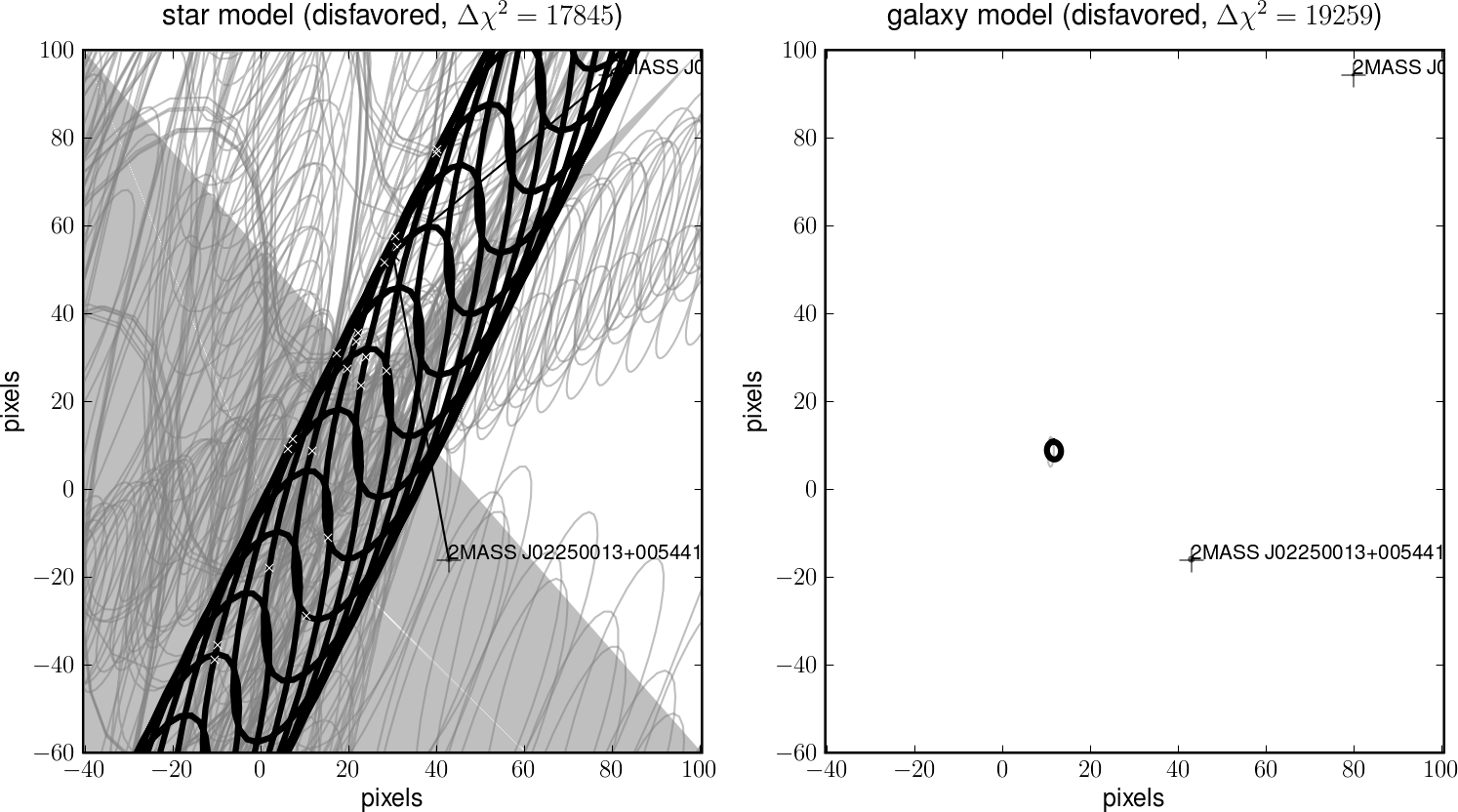}}\\[2ex]%
\resizebox{\examplefigurewidth}{!}{\includegraphics{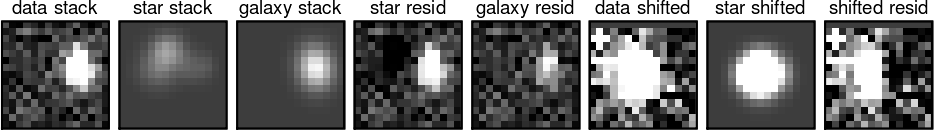}}
\caption{The same as \figurename~\ref{fig:example} but for a spurious
  source caused by a blinking artificial satellite.  Here the ``junk''
  model is favored over the point-source and galaxy models.  Note that
  the jackknife errors in the point-source and galaxy models are very
  large.  This source actually can be removed by flag-checking in the
  SDSSSS Coadd Catalog, but we show it here for illustrative
  purposes.\label{fig:examplecrap}}
\end{figure}

\clearpage
\begin{figure}
\includegraphics[width=\textwidth]{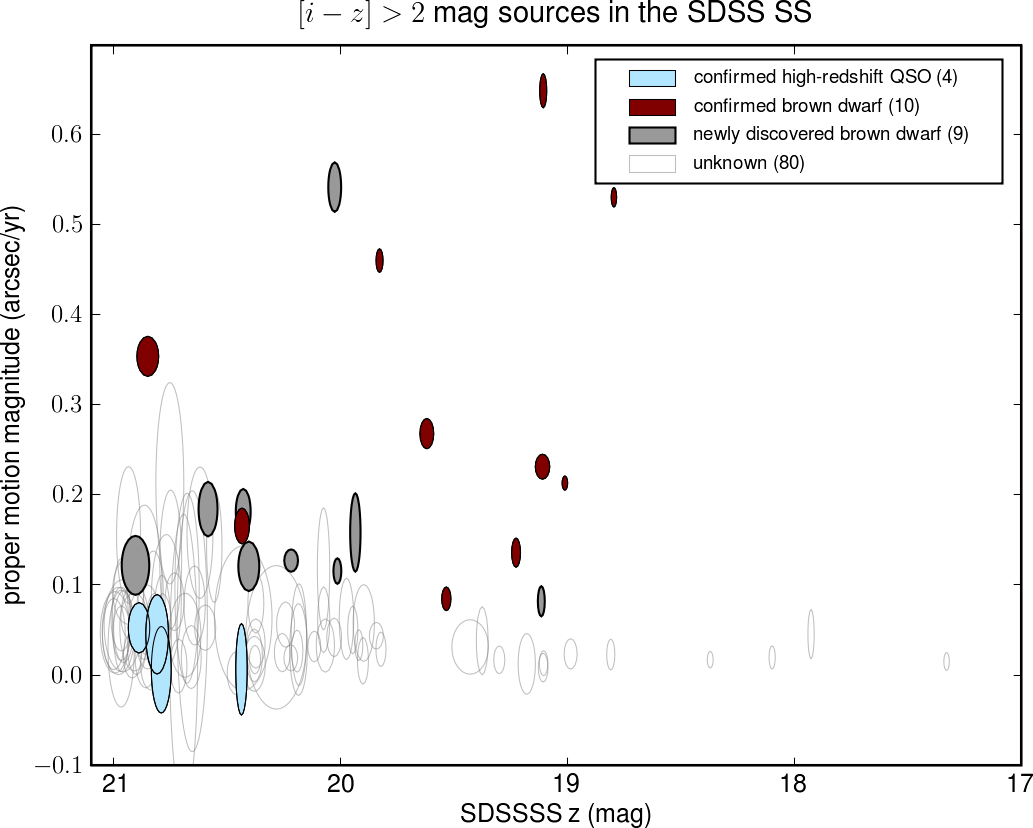}
\caption{Proper motion magnitude (angular speed) as a function of
  SDSSSS Coadd Catalog $z$-band magnitude for $[i-z]>2$~mag sources in
  the SDSS Southern Stripe that are preferentially described as point
  sources (by our $\chi^2$ hypothesis test).  The uncertainty regions
  are shown as transparent ellipses.  The spectroscopically confirmed
  high-redshift quasars \citep{jiang08a} and brown dwarfs
  \citep{fan00a, geballe02a, hawley02a, berriman03a, knapp04a,
    chiu08a, metchev08a} are shown in color.  Every one of the brown
  dwarfs has a significantly measured proper motion; none of the
  quasars do.  Other brown-dwarf candidates are clearly visible as
  significant movers (see also \tablename~\ref{tab:movingsources}).
  Note that the single-epoch detection limit is approximately
  $z=20.5~\mag$ in good seeing conditions.\label{fig:bubbles}}
\end{figure}

\clearpage
\begin{figure}
\includegraphics[width=\textwidth]{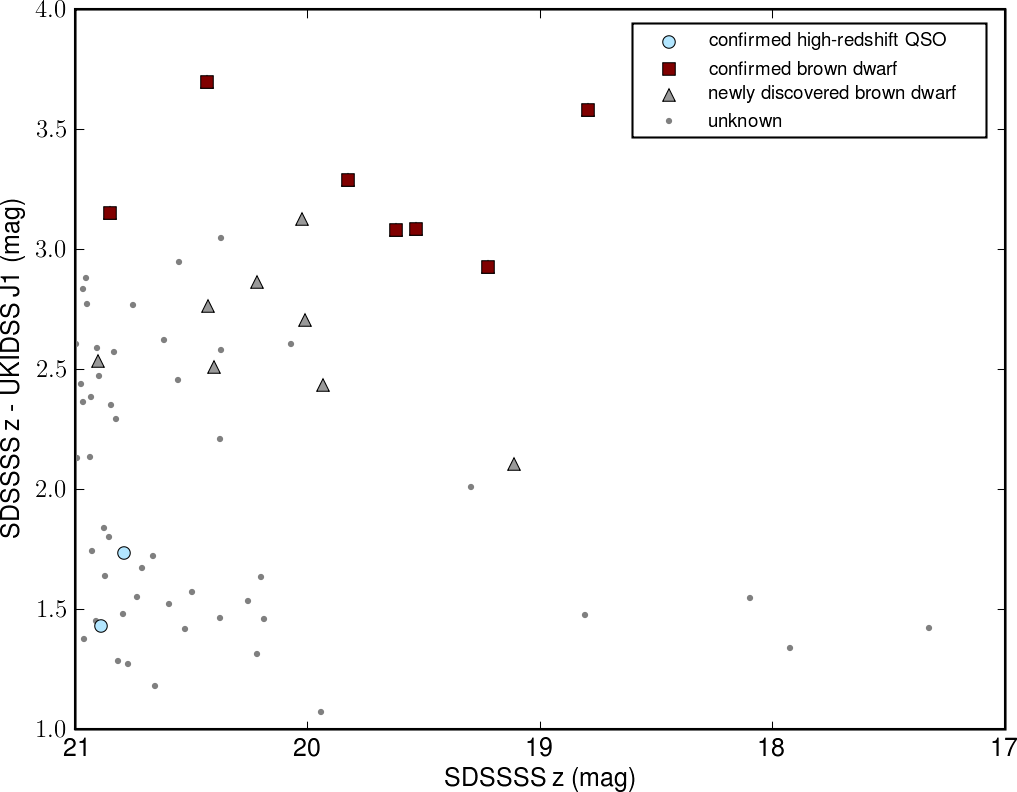}
\caption{UKIDSS and SDSSSS $[z-J]$ color plotted against SDSSSS
  $z$-band magnitude, for the sources in \figurename~\ref{fig:bubbles}
  that are detected in the UKIDSS $J$ band.  All of the
  significantly-moving objects have the very red $[z-J]$ colors of
  brown dwarfs; the likely brown dwarfs could have been identified by
  their proper motions and SDSSSS Coadd Catalog colors alone.  Note
  that the $z$ magnitude is in the AB system, while the $J$ magnitude
  is Vega-based.\label{fig:colormag}}
\end{figure}

\clearpage
\begin{figure}
\includegraphics[width=\textwidth]{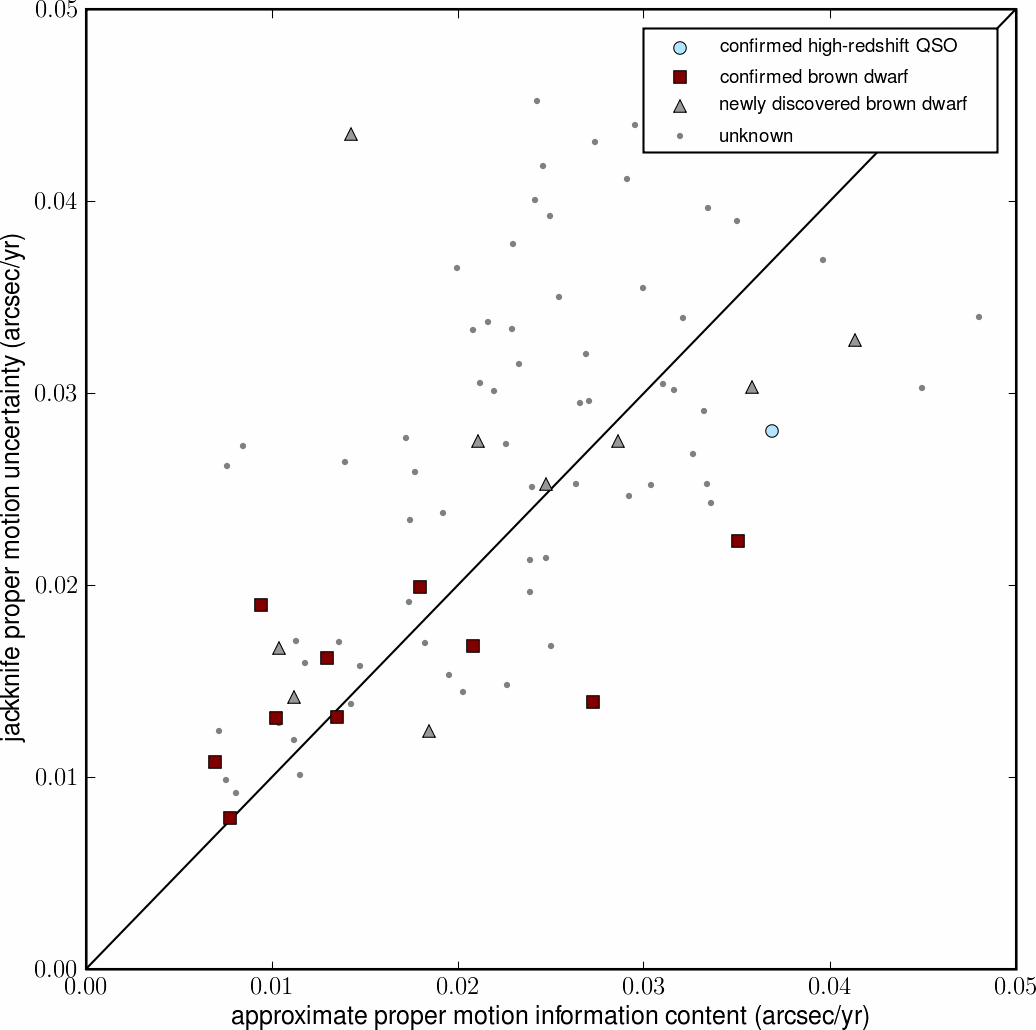}
\caption{Comparison of the jackknife-estimated proper-motion
  uncertainties to ``best-case'' values estimated from general
  principles of information in the imaging: The information estimate
  is the mean (square-signal-to-noise-weighted) imaging point-spread
  function FWHM divided by the signal-to-noise of the flux measurement
  (taken to be a proxy for the total detection signal-to-noise),
  divided by the root-variance of the time span.  This figure shows
  that the measurements are roughly as precise as they \emph{can be,}
  given the information content of the set of images; see
  \equationname~(\ref{eq:muerror}).\label{fig:info}}
\end{figure}

\clearpage
\begin{figure}
\includegraphics[width=\textwidth]{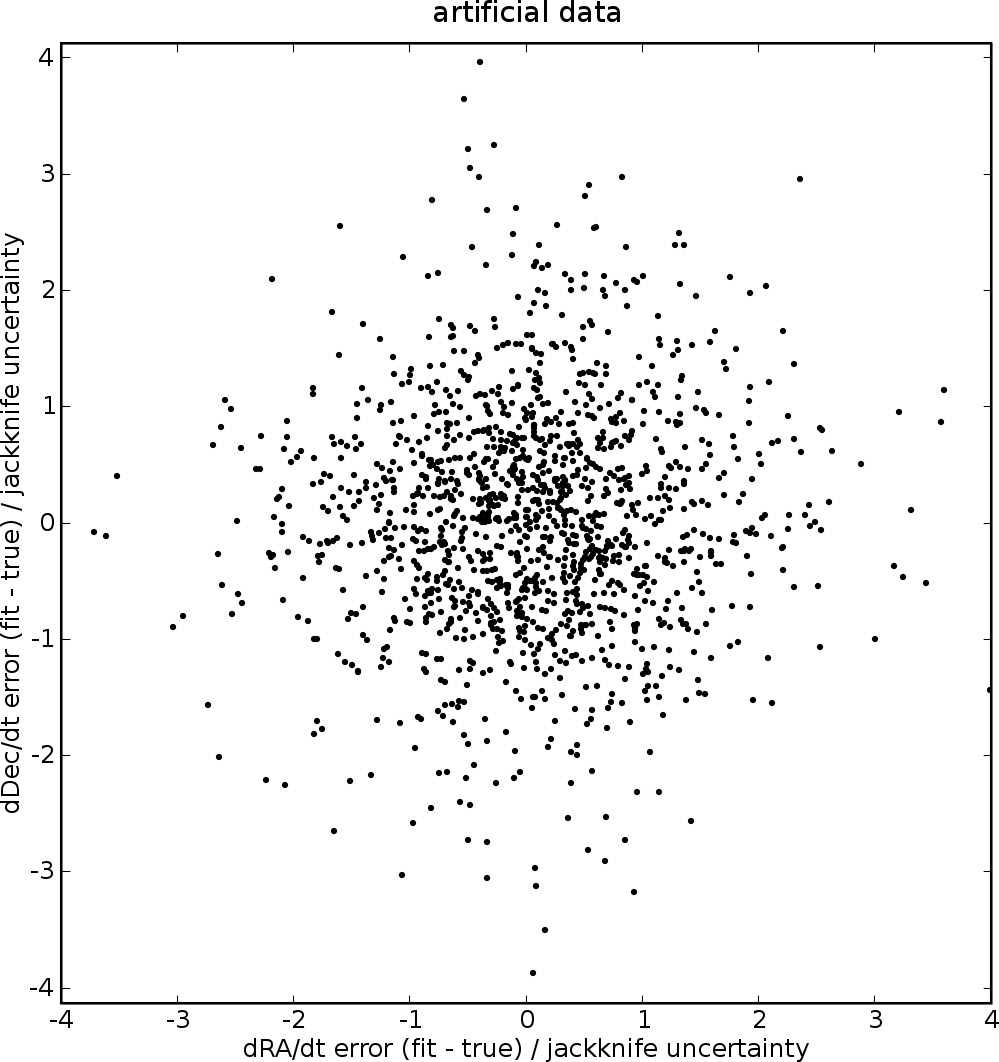}
\caption{Accuracy of jackknife uncertainty estimates for fits to
  artificial data made with known point-source properties.  Errors in
  fit parameters (fit minus true) have been divided by jackknife
  uncertainties.  The artificial data sets have identical imaging
  properties (noise amplitude, WCS, and PSF) to SDSS Southern Stripe
  sources, but contain artifical images made with point sources with
  true positions, fluxes and motions derived from the data as
  described in the text.  The cloud of points is centered near
  $(0,0)$, is circularly symmetric, and appears roughly Gaussian: our
  estimates are unbiased, uncorrelated, and have the expected
  magnitude and distribution of error.\label{fig:fake}}
\end{figure}

\end{document}